\documentclass{moriond}

\def\Journal#1#2#3#4{{#1} {\bf #2}, #3 (#4)}

\def\APJ{\em Astrophys. J.}
\def\APJL{{\em Astrophys. J.} Lett.}
\def\EPJ{\em EPJ Web of Conf.}
\def\JCAP{\em J. Cosmol. Astropart. Phys.}

\def\NIMA{{\em Nucl. Instrum. Methods} A}

\def\POS{\em Proc. of Science}

\def\UNIV{\em Universe}

\def\be{\begin{equation}}
\def\ee{\end{equation}}
\def\bea{\begin{eqnarray}}
\def\eea{\end{eqnarray}}

\usepackage{units,subfigure,amsmath,cleveref}

\newcommand{\Photo}{}

\begin{document}
\vspace*{4cm}
\title{Searches for primary photons and neutrinos with the Pierre Auger Observatory}

\author{Nicolás~M.~González$^*$ on behalf of the Pierre Auger Collaboration$^{**}$}

\address{$^*$Instituto de Tecnologías en Detección y Astropartículas (CNEA-CONICET-UNSAM),\\
Centro Atómico Constituyentes, Av. General Paz 1555, Buenos Aires, Argentina \\
$^{**}$Observatorio Pierre Auger, Av.\ San Mart{\'\i}n Norte 304, 5613 Malarg\"ue, Argentina\\
Full author list: \normalfont{\url{https://www.auger.org/archive/authors_2024_03.html}}}

\maketitle
\abstracts{
The Pierre Auger Observatory stands as the largest detector for ultra-high-energy (UHE) cosmic rays. The Observatory is also sensitive to UHE photons and neutrinos that can be produced along with UHE cosmic rays or in top-down processes, such as the decay of dark matter particles.
The search for these neutral particles relies on the hybrid measurements of extensive air showers, combining a fluorescence detector with a surface detector array and an underground muon detector. 
We present an overview of the searches for UHE photons and neutrinos utilizing data from the Pierre Auger Observatory. Currently, no photon or neutrino candidates have been identified. Consequently, we report on the most stringent limits to the integral UHE photon and neutrino fluxes above \unit[50]{PeV} and \unit[100]{PeV}, respectively, from diffuse and point-like steady sources. These limits led to strong constraints on theoretical models describing the cosmological evolution of the acceleration sites and the nature of dark-matter particles. Lastly, we briefly comment on the searches for these neutral particles in coincidence with gravitational wave events, underscoring the pivotal role of the Observatory in the context of multi-messenger astronomy at the highest energies.}

\section{Introduction}

Ultra-high energy (UHE) photons and neutrinos can be produced in cosmic-ray acceleration sites, during cosmic-ray propagation, or in the decay of dark-matter particles. 
While photons have a limited horizon due to the interaction with background radiation fields, neutrinos travel cosmological distances nearly unperturbed.
Thus, they are complementary messengers of astrophysical phenomena.

Primary astroparticles with UHEs interact with Earth's atmosphere initiating extensive air-showers that can be measured with ground-based detectors.
In this sense, the Pierre Auger is the largest cosmic-ray observatory in operation~\cite{Auger2015,AugerICRC2023}.
It has an outstanding exposure to photons and neutrinos above a few tens of PeV thanks to its instrumented area of \unit[$3000$]{$\text{km}^2$} and the measurements of air showers with several complementary detection systems~\cite{Roth2024}.
The surface detector (SD) is comprised of a main triangular array of $1660$ stations based on water-Cherenkov detectors spaced by \unit[1500]{m}. Two nested arrays of $61$ and $19$ stations with spacings of \unit[750]{m} and \unit[433]{m}, respectively, are employed to measure cosmic rays from below the second knee up to the ankle of the cosmic-ray energy spectrum. While the ground detectors are sensitive to secondary air-shower particles arriving on the ground, the fluorescence detector (FD), composed of $27$ telescopes across four sites overlooking the SD, is employed to measure the shower development in clear nights.
In addition, high-energy air-shower muons can be measured with buried scintillators forming the underground muon detector (UMD) in the two nested arrays.

In this contribution, we present a summary of the recent UHE photon and neutrino searches in the Pierre Auger Observatory.

\section{Searches for UHE photons}

Air showers initiated by primary photons are governed by electromagnetic (EM) processes.
They reach the maximum development at a deeper atmospheric depth, $X_\text{max}$, in comparison to their hadronic counterparts due to the lower multiplicity of EM interactions.
Photon-initiated showers also contain fewer secondary muons, as these are created in photonuclear interactions, which are highly suppressed compared to a hadron-initiated cosmic-ray shower. 
The lack of muons causes a steeper lateral fall-off and a slower rise of the signal registered by ground detectors than in the case of hadronic showers. These signatures of photon air showers are exploited to search for UHE photons in the Auger data.

Four discrimination methods, optimized in different energy ranges, are tailored and applied to the data. The measured observables are combined into a single discriminator using a multi-variate method. The median of the discriminator distribution for simulated photon events, leading to a $50\%$ signal efficiency, is called candidate cut. The number of measured and background events from simulations passing the candidate cut are compared to assess a possible photon flux detection.

Above \unit[50]{PeV}, the muon densities measured by the UMD stations of the \unit[433]{m} array are combined into a single event-wise observable~\cite{GonzalezICRC2023}. 
The relative background contamination ranges between $10^{-4}$ to $10^{-7}$ under the assumption of a pure-proton background, conservative in light of the deficit of muons identified in the air-shower simulations~\cite{Riehn2024}. 
The photon candidate cut was applied to \unit[$\sim$15]{months} of data, equivalent to an exposure of \unit[$\sim$0.6]{km$^2\cdot$sr$\cdot$yr}, and no photon candidate events were identified.

Above \unit[200]{PeV}, the measured $X_\text{max}$, the number of triggered SD stations of the \unit[750]{m} array, and a weighted sum of the integrated signals registered by them are combined using a boosted decision trees classifier~\cite{InfillPhotons2022}. 
Although a background contamination of two events was expected, no photon candidate events were spotted in the search data set spanning \unit[5.5]{yr} or \unit[$\sim$2.5]{km$^2\cdot$sr$\cdot$yr}.

Above \unit[1]{EeV}, the observables employed are $X_\text{max}$ and $F_\mu$, which is a proxy for the muon content developed within the air-shower Universality framework and based on the decomposition between EM and muonic components of the signals measured by the stations from the \unit[1500]{m} array~\cite{Savina2021}.
In the data set spanning \unit[12]{yr}, or \unit[$\sim$1000]{km$^2\cdot$sr$\cdot$yr}, $22$ photon candidate events were found with an expected background contamination of $30\pm15$.

Above \unit[10]{EeV}, the steeper lateral spread of particles and the longer risetime of the SD signals expected in photon showers are considered discrimination features~\cite{MainArrayPhotons2023}.
The observables are tuned with a subset of data, hence the method is free from assumptions about the cosmic-ray composition.
The searched data set spans \unit[16]{yr} with the \unit[1500]{m} array representing an exposure of \unit[$\sim$17,000]{km$^2\cdot$sr$\cdot$yr}.
Although $16$ candidate events are found, this number is compatible with the expected background contamination.

Given the absence of a photon signal in the Auger data, upper limits to the integral photon flux were established, as displayed in~\cref{fig:UL}, left, by circle markers~\cite{NiechciolICRC2023}.
These are the most stringent limits across over three decades in energy above \unit[50]{PeV}.
They can be used to constrain the mass, lifetime, and decay channels of dark matter particles, whose predicted cosmogenic fluxes are displayed as dashed lines~\cite{Universe2022}.
The cosmogenic fluxes from UHE cosmic-ray propagation are within reach above \unit[1]{EeV}. Conversely, the component due to UHE cosmic rays irradiating the Galaxy, which may be predominant below \unit[0.1]{EeV}, is around two orders of magnitude below the current sensitivity.

\section{Searches for UHE neutrinos}

The identification of UHE neutrinos is based on the search for very inclined showers with a prominent EM component, which is not expected for cosmic-ray showers due to the atmospheric attenuation~\cite{EeVNeutrinos2019}.
This feature is exploited by two detection channels.
The down-going (DG) channel is sensitive to air showers initiated by any neutrino flavor interacting deep in the atmosphere with zenith angles between \unit[60]{$^\circ$} and \unit[90]{$^\circ$}.
The earth-skimming (ES) channel is sensitive to $\tau$ neutrinos interacting in the Earth's mantle, producing a $\tau$ lepton that decays and initiates an upwards-going shower with zenith angles between \unit[90]{$^\circ$} and \unit[95]{$^\circ$}.
The Area-over-Peak (AoP) in an SD station is defined as the ratio of the integral signal to its peak value and normalized to the average signal produced by a single muon.
In the case of the ES channel, the discriminator is the average event-wise AoP, while a Fisher discriminant combining the AoP values of several stations is used for the DG channel.
An inclined cosmic-ray shower is dominated by muons, which are seen in the SD traces as narrow peaks, thus leading to AoP values closer to unity. In the case of neutrinos, the traces have a rich structure driven by the EM component, leading to AoP values that are larger than unity.

The neutrino candidate cut is set such that one background event is expected to pass it in $50$ (\unit[20]{yr}) for the ES (DG) channel.
At energies below \unit[$\sim$40]{EeV}, the exposure is dominated by the ES channel, while at higher energies, the DG kicks in because of the lower probability of detecting the upwards-going shower from the $\tau$ decay~\cite{NiechciolUHECR2023}. The total exposure grows from \unit[$10^{16}$]{cm$^2\cdot$s$\cdot$sr} at \unit[0.1]{EeV} to \unit[$10^{18}$]{cm$^2\cdot$s$\cdot$sr} at \unit[100]{EeV} in the searched data set spanning around \unit[18]{yr}.
The energy-integrated sensitivity is dominated by the ES channel ($79\%$), while the overall flavor sensitivity is maximum for $\tau$ neutrinos ($86\%$).

\begin{figure}[!tb]
\centering
\includegraphics[width=0.47\textwidth]{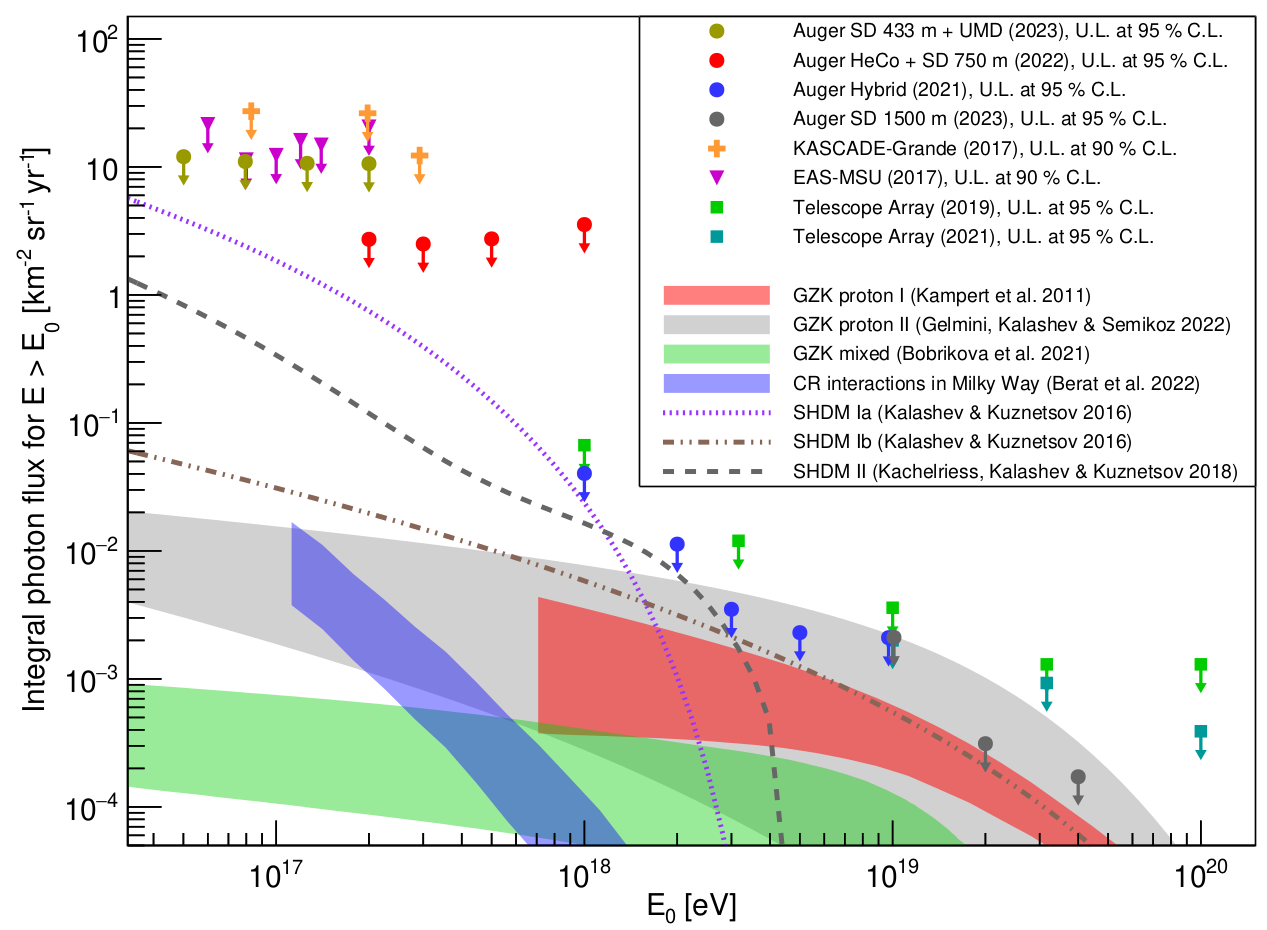}\includegraphics[width=0.47\textwidth]{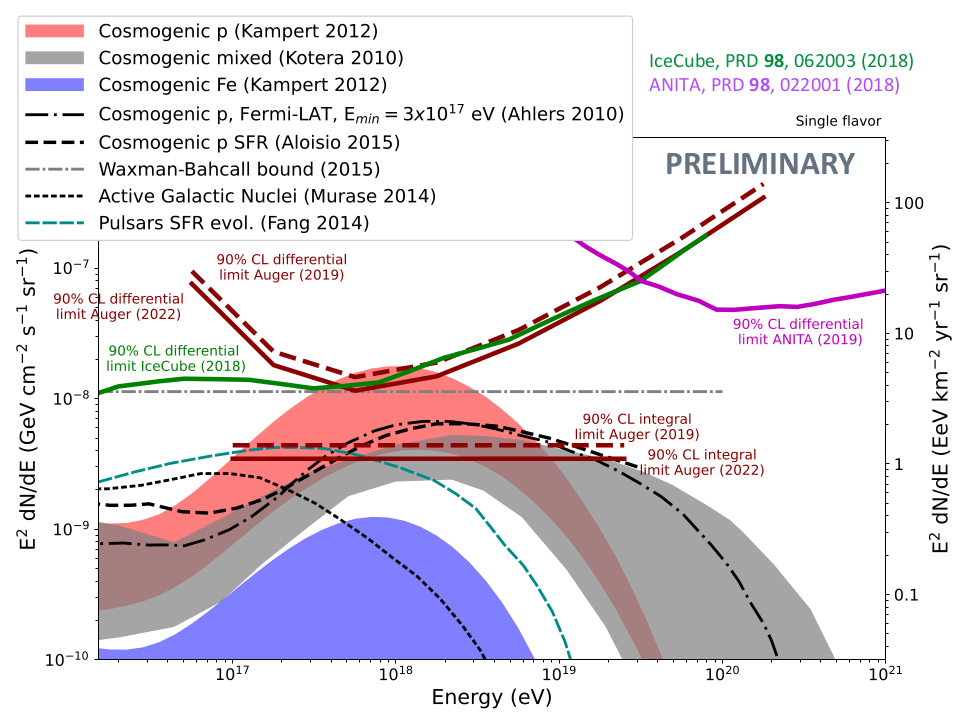}
\caption{Left: The upper limits on the integral UHE photon flux established with Auger data (circles)~\protect\cite{NiechciolICRC2023}. Triangles and squares represent previous results of various experiments. Filled bands denote the expected cosmogenic photon fluxes assuming different cosmic-ray compositions. The predicted fluxes from the decay of dark-matter particles are shown as dashed lines. Right: The upper limits at $90\%$ confidence level on the diffuse and integral UHE neutrino flux obtained with Auger data~\protect\cite{NiechciolUHECR2023} (dark red lines), accompanied by those estimated with data from other observatories and predicted by several cosmogenic and astrophysical models.}
\label{fig:UL}
\end{figure}

No events in our data passed the candidate cuts. Thus, differential upper limits to the diffuse neutrino flux, shown in solid dark red in~\cref{fig:UL}, right, are established.
The best sensitivity is achieved at around \unit[1]{EeV} and is comparable to that of IceCube.
The integral upper limit of \unit[$3.5\times10^{-9}$]{GeV$\cdot$cm$^{-2}$s$^{-1}$sr$^{-1}$} in the energy range between $0.1$ and \unit[25]{EeV}, corresponding to $90\%$ of the events, has been employed to scan the parameter space of the cosmic-ray sources~\cite{Guido2024}.
The assumption of a pure-proton injection at the sources is strongly constrained, whereas more loose conditions can be computed for a mixed-composition~\cite{PetrucciICRC2023}.

Due to Earth’s rotation and the location of the Observatory, a point source at a given declination within \unit[$-$85]{$^\circ$} and \unit[60]{$^\circ$} transits through the field-of-view of the ES (DG) channel at most during \unit[4]{h} (\unit[11]{h}) per day.
Hence, the sensitivity to a neutrino flux from a point source depends on its declination~\cite{NeutrinoPointSources2019}.
The maximum sensitivity is attained for declinations $\pm55^\circ$, which corresponds to the longest transit of a point source in the ES channel.
The point-source sensitivity is relevant for the follow-up searches after the detection of significant astrophysical events.
For instance, after the detection of the gravitational wave (GW) event GW170817 by LIGO/VIRGO corresponding to a neutron star merger, a neutrino search at UHE was conducted with Auger data.
Since that event was transitting the ES field-of-view, a very high sensitivity was attained in a time window of \unit[500]{s} around the event timestamp, leading to the most stringent upper limits to the neutrino emission above \unit[0.1]{EeV}~\cite{GW1708172017}.
Another example of a follow-up neutrino search is the one related to the blazar TXS0506+056, later identified as a neutrino source by IceCube~\cite{TXS0506+0562020}.
One event at EeV energies would have been measured by Auger if the extrapolated spectrum were sufficiently harder than the measured one by IceCube.
 
Multi-messenger analyses stacking several merger events detected by LIGO/VIRGO have also been carried out.
The exposures to a possible neutrino signal coming from $83$ binary black hole mergers have been combined to compute an upper limit to the daily neutrino luminosity of \unit[$2.3\times10^{53}$]{erg} from this kind of phenomena assuming a constant emission with a $E^{-2}$ spectrum~\cite{SchimpICRC2021}.
Similarly, a first search for photons above \unit[10]{EeV} from $10$ nearby or well-localized GW events has been carried out~\cite{PhotonsGWs2023}. Upper limits to the daily spectral fluence ranging between $30$ and \unit[40]{MeV$\cdot$cm$^{-2}$} were established for each source. Particularly, an upper limit of $\sim20\%$ on the energy transferred to primary photons above \unit[40]{EeV} was set from the nearby merger GW170817.

\section{Summary}

The Pierre Auger Observatory has an unrivaled exposure to neutral primaries above \unit[50]{PeV}. Its data has been employed to set the most stringent upper limits to the diffuse photon flux across three decades in energy. 
A background-free sensitivity to diffuse neutrinos and, depending on the declination, also to point sources has been attained above \unit[0.1]{EeV}.
Thanks to its large sky coverage, Auger provides valuable information at UHE in the context of multi-messenger astronomy, which will be further improved with new data acquired with the upgraded observatory codenamed AugerPrime~\cite{Suomijarvi2024}.

\end{document}